\documentclass[pra,superscriptaddress,showpacs,twocolumn]{revtex4}
\usepackage{graphicx}       
\begin{document}
\title{Joint measurements of spin, operational locality and uncertainty}
\author{Erika Andersson}
\affiliation{Department of Physics, University
of Strathclyde, Glasgow G4 0NG, UK}
\author{Stephen M. Barnett}
\affiliation{Department of Physics, University
of Strathclyde, Glasgow G4 0NG, UK}
\author{Alain Aspect}
\affiliation{Department of Physics, University
of Strathclyde, Glasgow G4 0NG, UK}
\affiliation{Groupe d'Optique Atomique, Laboratoire Charles Fabry de l'Institut d'Optique, UMR 8501 du CNRS, 91403 Orsay CEDEX, France}
\date{\today}
\begin{abstract}
Joint, or simultaneous, measurements of non-commuting observables {\it are} possible within quantum mechanics, if one accepts an increase in the variances of the jointly measured observables. In this paper, we discuss  joint measurements of a spin 1/2 particle along any two directions. Starting from an operational locality principle, it is shown how to obtain a  bound on how sharp the joint measurement can be. We give a direct interpretation of this bound in terms of an uncertainty relation.
\end{abstract}
 

\maketitle

\section{Introduction}
Quantum mechanics places restrictions on how sharply two non-commuting observables can be measured jointly. A joint measurement means that by performing {\it one} measurement on a {\it single} quantum system, we are able to produce a result for {\it each} of the two observables. 
This could, just to take an example, be useful when trying to eavesdrop on two parties, who are communicating using quantum cryptography. Let us assume that the protocol used is BB84 \cite{BB84}, where two different photon polarisation bases are used to represent the data sent. A polarised photon is of course equivalent to a spin 1/2 quantum system. In the first basis, horizontal polarisation means ``0" and vertical polarisation means ``1". The  second basis is oriented at 45$^\circ$ to the first. You are trying to measure whether a ``0" or ``1" is sent, but you do not know what basis the sender is using for each photon. In this situation, you might try to measure polarisation along {\it both} directions at the same time, so that, when the sender announces which basis was used, you can pick the right result. It turns out, of course, that an eavesdropper making quantum mechanical joint measurements of polarisation along two non-orthogonal directions, will not obtain perfect information about the polarisation, even after the bases have been announced. Nevertheless, the example illustrates that it may be interesting to consider joint measurements in quantum mechanics, and that it is important to understand the limitations placed on such measurements.

One way to achieve a joint measurement of two observables would simply be to measure one of the observables, and to guess a result for the other observable. This is not usually the best way to perform the joint measurement. It is also possible to make more balanced joint measurements, where the ``element of guessing" is distributed more evenly between the observables. The variances for jointly measured non-commuting observables, as constructed by considering the same joint measurement performed many times on an ensemble of identically prepared quantum systems, have to be larger than if one would make sharp measurements of the observables alone.  Quantum mechanical joint measurements have been considered, for example, in seminal papers by Arthurs and Kelly \cite{artkel} and Arthurs and Goodman \cite{artgood} and are reviewed in \cite{stig}. 

In this paper, we will consider joint measurements of a non-commuting pair of components of polarisation, or spin 1/2, along any two directions. A bound on the sharpness for such a measurement 
can be derived using the formalism of generalised measurements or POM (or POVM) measurements \cite{busch}. Here, we will show how to obtain the same bound without any explicit description of the measurement operators. We only use the assumption that a joint probability distribution for the measurement exists, together with a requirement for operational locality. By operational locality, we mean that if two quantum systems are space-like separated, then what is done to one of the systems locally cannot affect the reduced density matrix of the other system \cite{ghirardi}. This could also be referred to as a requirement that no (superluminal) signalling can take place. Furthermore, a joint probability distribution for the measured components clearly must exist for a joint measurement, whether the measurement is quantum or classical. These two assumptions, operational locality and the existence of a joint probability distribution, are enough to obtain the measurement bound.

We also show how the bound on the joint measurement may be written as an uncertainty relation for the increases in the measured variances. In this form, it is clearly seen how the polarisation or spin measurement bound fits in with the general result for joint measurements as stated by Arthurs and Goodman \cite{artgood}. The general result holds for any two observables, but the bound is not always tight for any measured state. The uncertainty relation for two jointly measured spin observables, on the other hand, can always be saturated with a suitable measurement, for any measured state.

The paper is organised as follows. In section \ref{secsim}, we introduce the joint spin measurement and derive a bound on its sharpness using a locality argument. In order to illustrate the concept of a joint spin measurement, we present an example of a possible realisation in section \ref{secjointmeas} along with the relevant measurement operators in section \ref{secjointop}. In section \ref{secunc}, the bound is shown to be equivalent to a bound on the product of the increases in the variances of the jointly measured observables. Finally a discussion and conclusions are offered.

\section{Joint measurements of spin}
\label{secsim}
Let us suppose that we measure the spin of a $S=1/2$ particle jointly along two directions, given by the unit vectors ${\bf a}$ and ${\bf a'}$. At the moment, we do not need to think about how to achieve the joint measurement. If measured separately, the relevant observables are given by ${\bf\hat{A}}={\bf a\cdot\hat{\sigma}}$ and ${\bf\hat{A}'}={\bf a'\cdot\hat{\sigma}}$. The question now arises what requirement to place on the joint measurement, in order for it to be a ``good" measurement of both observables. A frequently made choice is the joint unbiasedness condition employed e.g. by Arthurs and Goodman \cite{artgood}. In line with this condition, we choose to require that the constructed expectation values of the jointly measured observables must be proportional to the expectation values of the separately measured observables.  
This condition must hold for any measured quantum state with the same constants of proportionality, called $\alpha$ and $\alpha '$.
Using this fact, the variances of the jointly measured observables may be written as
\begin{eqnarray}
\label{simvar}
(\Delta{{A}_J})^2&=&\overline{{A}_J^2} - \overline{{A}_J}^2 = 1- \alpha^2\langle{\bf\hat{A}}\rangle^2\nonumber\\
(\Delta{{A}'_J})^2&=&\overline{{A}_J^{\prime 2}} - \overline{{A}_J'}^2 = 1- \alpha^{\prime 2}\langle{\bf\hat{A}'}\rangle^2,
\end{eqnarray}
where we denote the values obtained in the joint measurement by ${A}_J$ and ${A}'_J$.  An overbar denotes an average, taken for the state we are measuring. We choose not to use the notation $< .  >$ used for quantum mechanical averages at this point, since we have defined neither observables nor operators for the joint measurement. $A_J$ and $A'_J$ are results obtained in the measurement, not operators. 
In equation (\ref{simvar}), we have also used the fact that the measurement result, $\pm 1$, always equals +1 when squared.  In general, the joint measurement of $\bf\hat{A}$ and $\bf\hat{A}'$ results in an increase in their variances as compared to  separate measurements, and this forces $|\alpha|$ and $|\alpha '|$ to be smaller than 1. The precise upper bound on  $|\alpha|$ and $|\alpha '|$ stems from the fact that a joint probability distribution must exist for  ${A}_J$ and ${A}'_J$, for any valid quantum state. The  bound will depend on the directions of $\bf a$ and $\bf a'$. It has previously been derived by considering all possible generalised measurement operators  describing the joint measurement \cite{busch}. In the following, we present a less technical derivation using the principle of operational locality. This requires no further assumptions about the joint measurement itself, other than the definitions made above.

We now proceed to derive a bound on the joint measurement. Consider two spin-1/2 particles prepared in the singlet state
\begin{equation}
|\psi^-\rangle = {1\over\sqrt{2}}\left(|+\rangle _1|-\rangle _2 - |-\rangle _1|+\rangle _2\right).
\end{equation}
Two observers have access to one quantum system each. By operational locality, we mean that no operation done on one of the systems can affect the reduced density matrix of the other system. The local operation can be a measurement, or any other operation. Local operations on one system thus cannot be detected on the other. Measurement results on the two subsystems may be correlated, but the correlations cannot be used for signalling \cite{ghirardi}. It necessarily follows that the communication scheme we now will describe must  fail. 
On quantum system 2, observer 2 will make a measurement of spin {\it either} along ${\bf b}$ or along ${\bf b}'$. This yields the results $\pm 1$ with equal probabilities.
On quantum system 1, observer 1 will then make a joint measurement of spin along two directions, ${\bf a}$ and ${\bf a}'$. 
Consider the situation when ${\bf b}$ is parallel to ${\bf a} + {\bf a}'$ and ${\bf b}'$ is parallel to ${\bf a} - {\bf a}'$. Intuitively, if observer 2 chooses to measure along ${\bf b}$, observer 1 should be likely to obtain the same result for both ${\bf a}$ and ${\bf a}'$, ``++" or ``$- -$", and different results, ``$+-$" or ``$-+$",  if observer 2 measures along ${\bf b}'$. If so, this would provide a means for instantaneous communication between the two observers. 
But because of operational locality, the probabilities for the results observer 1 obtains cannot depend on any action taken by observer 2. Observer 1 cannot tell whether observer 2 measured ${\bf b\cdot\hat{\sigma}}_2$ or ${\bf b'\cdot\hat{\sigma}}_2$, and the communication scheme has to fail. This  will provide a bound on how accurately observer 1 can perform the joint measurement.

Let us denote the measurement results by $A_J$, $A'_J$, $B$ and $B'$; these are all $\pm 1$. Because of the operational locality principle, the probabilities for observer 1 to obtain the same result for spin along both $\bf a$ and $\bf a'$ cannot depend on whether observer 2 measured along $\bf b$ or along $\bf b'$.
Nevertheless, suppose first that observer 2 has measured spin along $\bf b$. The probability that observer 1 obtains $A_J=A'_J$  can then be written
\begin{equation}
p(A_J=A'_J)=p(A_J=A'_J=B)+p(A_J=A'_J=-B).
\end{equation}
The probabilities on the right hand side, for the triples $A_J, A'_J, B$ and $A_J, A'_J, B'$, must exist. These probabilities are greater than or equal to zero and hence
\begin{eqnarray}
\label{ineq}
&&p(A_J=A'_J=B)+p(A_J=A'_J=-B) \nonumber\\ 
&&\geq |p(A_J=A'_J=B)-p(A_J=A'_J=-B)|.
\end{eqnarray}
We can use the correlation functions
\begin{equation}
E(A,B)=p(A=B)-p(A=-B)=\overline{AB}
\end{equation}
to write 
\begin{eqnarray}
p(A_J=A'_J=B)-p(A_J=A'_J=-B) \nonumber \\ ={1\over 2}\left[E(A_J,B)+E(A'_J,B)\right],
\end{eqnarray}
finally giving us
\begin{equation}
p(A_J=A'_J)\geq {1\over 2}|E(A_J,B)+E(A'_J,B)|.
\end{equation}
In a similar way, if  we assume that observer 2 has measured spin along $\bf b'$, we can derive
\begin{equation}
p(A_J=-A'_J)\geq{1\over 2}|E(A_J,B')-E(A'_J,B')|.
\end{equation}
The probabilities on the left hand sides of these two inequalities are independent of whether observer 2 measured spin along $\bf b$ or $\bf b'$. Adding the two inequalities, and noting that $p(A_J=A'_J)+p(A_J=-A'_J)=1$, we obtain
\begin{equation}
\label{belleq}
|E(A_J,B)+E(A'_J,B)|+|E(A_J,B')-E(A'_J,B')|\leq 2.
\end{equation}
This inequality bears great resemblance to a CHSH Bell inequality \cite{bell, chsh}. In the context of Bell inequalities, it has been shown that the existence of ``hidden variables" reproducing the correct probability distributions  is equivalent to the condition that joint probabilities exist for all triples \cite{fine}.
Quantum mechanical violations of Bell inequalities mean that either hidden variables cannot exist, or quantum mechanics has to be nonlocal. In the present case, on the other hand, the inequality (\ref{belleq}) {\it must be satisfied} for joint measurements in quantum mechanics \cite{demuynck}. This is because joint probability distributions must necessarily exist for jointly measured observables, whether the measurement is quantum mechanical or not, and so Bell's inequality must be valid for a joint measurement. This argument, however, does not tell us why the Bell inequality actually will give a {\it tight} condition in our particular case of joint measurement. In principle, the condition that a joint probability distribution is compatible with a quantum mechanical joint measurement, is stronger than the condition that it should merely exist as a classical probability distribution. As an example, the joint probability distribution $p_{++}=1, p_{+-}=p_{-+}=p_{--}=0$ is possible classically, but is not attainable as a quantum mechanical probability distribution for a joint measurement of two non-commuting observables.

\begin{figure}
\center{\includegraphics[width=7cm,height=!]{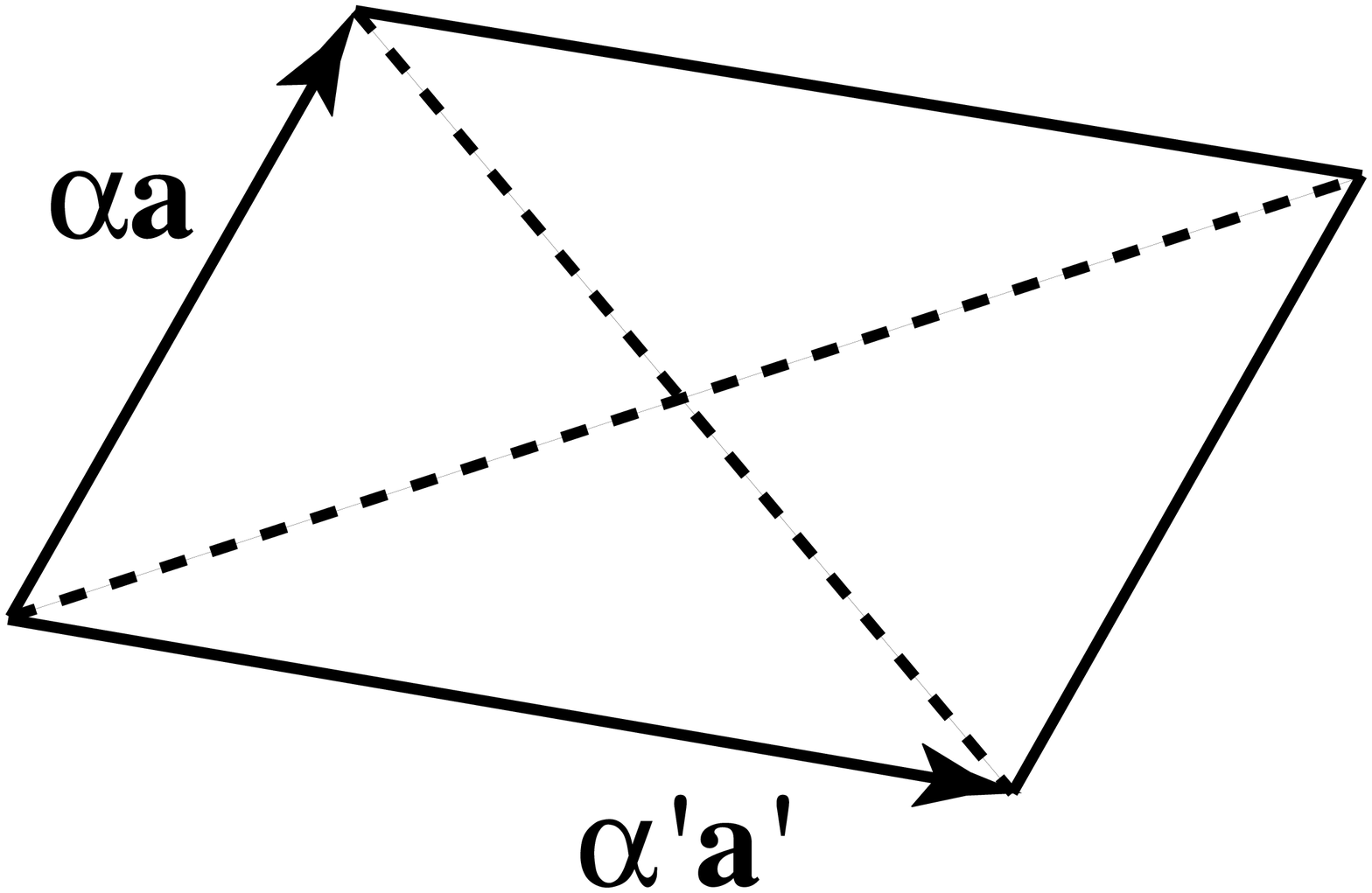}}
\caption{For a joint measurement of spin along both $\bf a$ and $\bf a'$ to be possible, the sum of the diagonals in a parallellogram with $\bf \alpha a$ and $\bf  \alpha 'a'$ as its sides must be less than 2. Unless $\bf a$ and $\bf a'$ are parallel, this forces both $|\alpha|$ and $|\alpha '|$ to be strictly less than 1.}
\label{parallelogram}
\end{figure}

Inequality (\ref{belleq}) places restrictions on the correlations between observables of the two quantum systems. We would like to obtain a quantum mechanical bound on the joint measurement of $\bf\hat{A}$ and $\bf\hat{A}'$, involving only observer 1 and quantum system 1. If spin is measured only along $\bf a$ on quantum system 1, and along $\bf b$ on quantum system 2, the quantum mechanical correlation function is given by
\begin{equation}
E(A,B)= \langle\psi^-|{\bf a}\cdot\hat{\sigma}_1 {\bf b}\cdot\hat{\sigma}_2|\psi^-\rangle = -{\bf a\cdot b}.
\end{equation}
Since joint measurements reduce expectation values by factors $\alpha$ and $\alpha '$ for any state, we must have 
\begin{equation}
E(A_J,B)= \alpha\langle\psi^-|{\bf a}\cdot\hat{\sigma}_1 {\bf b}\cdot\hat{\sigma}_2|\psi^-\rangle = -\alpha {\bf a\cdot b},
\end{equation}
and similarly for $E(A'_J,B), E(A_J,B')$ and $E(A'_J,B')$. Using this in (\ref{belleq}) gives
\begin{equation}
|-(\alpha{\bf a} + \alpha '{\bf a}')\cdot {\bf b}|+|-(\alpha{\bf a} - \alpha '{\bf a}')\cdot {\bf b}'|\leq 2.
\end{equation}
This must be valid for any choice of $\bf b$ and $\bf b'$. The left hand side is maximised when $\bf b$ is chosen parallel to $\bf \alpha a + \alpha ' a'$ and $\bf b'$ parallel to $\bf \alpha a - \alpha ' a'$, giving
\begin{equation}
\label{alphabound}
{\bf |\alpha a + \alpha ' a'| + |\alpha a - \alpha ' a'|}\leq 2.
\end{equation}
This condition, linking $\alpha, \alpha ', \bf a$ and $\bf a'$, is the same as obtained in \cite{busch}.  This inequality has a simple geometrical meaning as illustrated in figure \ref{parallelogram}:  the sum of the lengths of the diagonals in a parallellogram with $\bf \alpha a$ and $\bf \alpha' a'$
as its sides must be less than 2. Unless the unit vectors $\bf a$ and $\bf a'$ are parallel, this forces both $|\alpha|$ and $|\alpha '|$ to be strictly less than 1. The smaller $|\alpha|$ and $|\alpha '|$ are, the more smeared the jointly measured observables are, since this increases their variances according to equations (\ref{simvar}). In \cite{bushill}, a situation corresponding to $\bf a$ and $\bf a'$ being orthogonal to each other arises.

The derivation in \cite{busch} was made by explicitly considering the possible generalised measurement operators describing the joint measurement, whereas the present derivation is based on the assumption that joint probability distributions exist for the two measured spin components, and on the principle of operational locality. The locality principle is used in much the same way as the energy conservation principle may be used to solve physical problems. It is not the only way to arrive at the conclusion, but may simplify calculations considerably. In this case, the advantage is that the description of the joint measurement can be left open, and the derivation is not tied to any particular model of joint quantum measurements.

An interesting connection to Cirel'son's inequality \cite{cirel} can also be made. Cirel'son showed that quantum systems will obey the less stringent bound
\begin{equation}
\label{cireleq}
|E(A,B)+E(A',B)|+|E(A,B')-E(A',B')|\leq 2\sqrt{2}.
\end{equation}
Translated into a bound on a measurement on system 1, this would give
\begin{equation}
\label{cirelalpha}
{\bf |\alpha a + \alpha ' a'| + |\alpha a - \alpha ' a'|}\leq 2\sqrt{2}.
\end{equation}
This condition does not restrict $|\alpha|$ and $|\alpha'|$ to be smaller than one. The reason for this is clear. If we do not require to make a joint measurement, sharp measurements of each observable are possible. Cirel'son's inequality, which is satisfied by correlations in a quantum system, does not lead to any restriction on the sharpness of the measurement. 

\section{An example}
\label{secjointmeas}

Let us now consider how to realise a joint measurement of $\bf\hat{A}$ and $\bf\hat{A}'$, satisfying the bound (\ref{alphabound}). Our fundamental requirement is that the joint measurement, performed on a single spin 1/2 system, gives us {\it four} possible results, $++, +-, -+$ or $--$. There are many ways to achieve this, but we will look at one particular method. Suppose that we perform a measurement of {\it one} of the two spin components $\bf c\cdot\hat{\sigma}$ and $\bf c'\cdot\hat{\sigma}$, with probabilities $p$ and $1-p$ respectively, obtaining a result $C$ or $C'$. 
\begin{figure}
\center{\includegraphics[width=8cm,height=!]{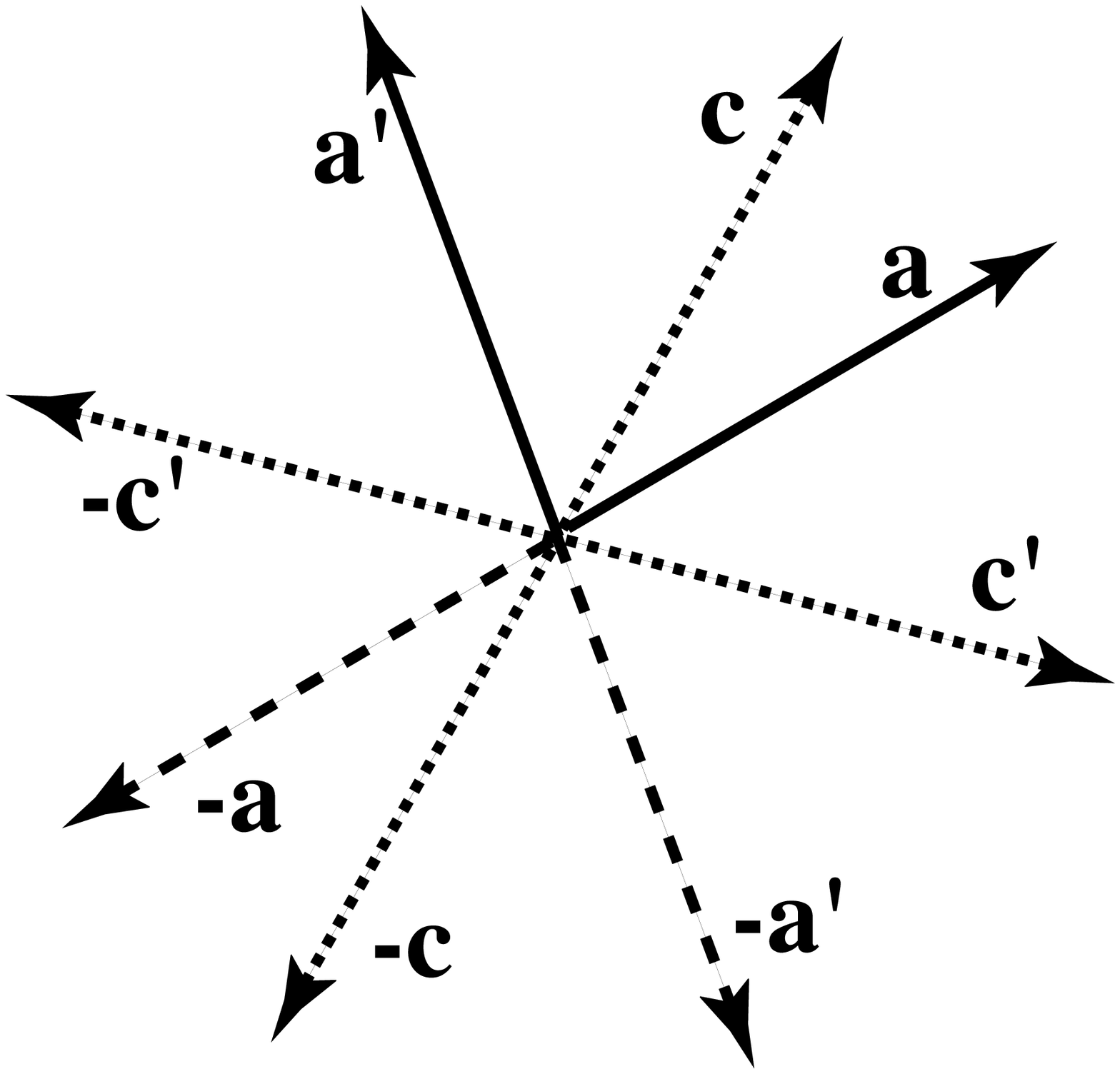}}
\caption{We are proposing to make a joint measurement of spin along directions $\bf a$ and $\bf a'$ by making a measurement along {\it either} direction $\bf c$ {\it or} direction $\bf c'$. Direction $\bf c$ will lie somewhere between $\bf a$ and $\bf a'$. Direction $\bf c'$ will lie somewhere between $\bf a$ and $\bf -a'$. In this picture, orthogonal spin states $|+_a\rangle$ and $|-_a\rangle$ correspond to opposite directions $\bf a$ and $\bf -a$, and similarly for the other directions. If we want to think in terms of photon polarisation, the orthogonal polarisation states  $|H\rangle$ and $|V\rangle$ will also be represented by opposite vectors in the picture. The vectors $\bf a$, $\bf a'$ etc. are the Bloch vectors for the corresponding polarisation states.}
\label{spindirections}
\end{figure} 
We then try to associate this with a joint measurement of $\bf a\cdot\hat{\sigma}$ and $\bf a'\cdot\hat{\sigma}$ in the following way. If we choose to measure along $\bf c$, and obtain $C=+1$, then we say that the result is $A_J=A'_J=+1$, and if $C=-1$, then we say that $A_J=A'_J=-1$. 
In a related way, in \cite{bushill}, the marginal distributions of a four-outcome measurement are used to construct unsharp measurements along certain directions.
Intuitively, direction $\bf c$ should lie perhaps not exactly halfway, but somewhere between $\bf a$ and $\bf a'$, as in figure \ref{spindirections}. 

\begin{figure}
\center{\includegraphics[width=10cm,height=!]{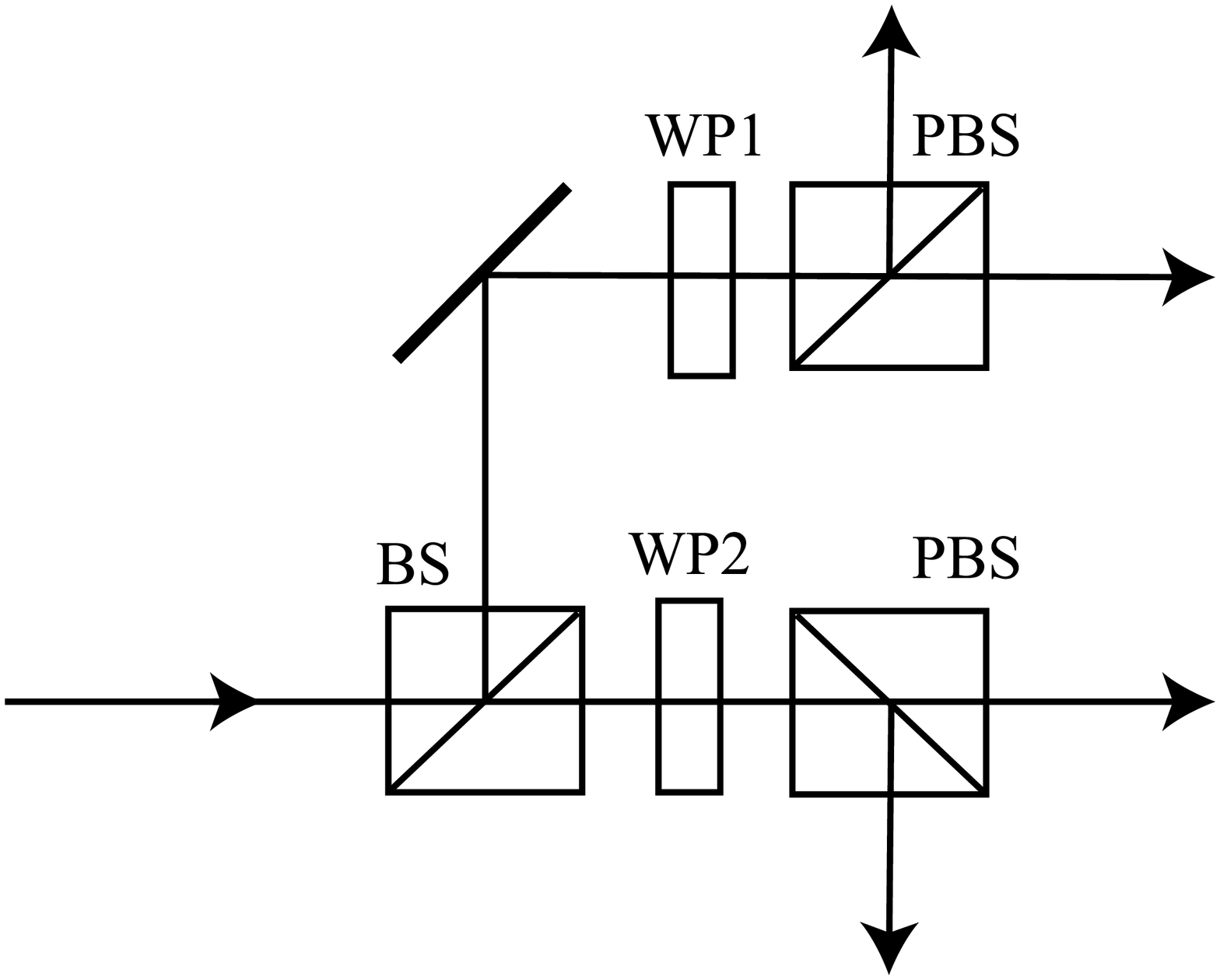}}
\caption{One way to realise a measurement of polarisation {\it either} along one direction {\it or} along another, would be to use a non-polarising beam splitter (BS) with a suitable splitting ratio, and to measure polarisation along the first direction in one arm, and along the second direction in the other arm. The wave plates WP1 and WP2 rotate the polarisation in the beams in a suitable way so that the polarisation measurements  may be done using polarising beam splitters PBS, which separate horizontal and vertical polarisation. The detectors are not shown in the picture. With this setup, there are {\it four} possible measurement outcomes, even for a single photon.}
\label{bsrealisation}
\end{figure}

If we measure along $\bf c'$, then, if $C'=+1$, we say that the result is $A_J=+1$ and  $A'_J=-1$, and if $C'=-1$, then we say that $A_J=-1$ and $A'_J=+1$. 
Direction $\bf c'$ will lie somewhere between $\bf a$ and $\bf -a'$. One way to choose whether to measure along $\bf c$ or $\bf c'$, especially if we are measuring the polarisation of a photon, would be to use a beam splitter with the required splitting ratio, and then to measure along $\bf c$ in one output beam and along $\bf c'$ in the other, as in figure \ref{bsrealisation}. Such a setup has been considered for measurement along two orthogonal polarisation directions in \cite{buschbeam, demuynckbeam, novel}. The probabilistic choice between $\bf c$ and $\bf c'$ can, however, equally well be made entirely classically.

It is not self-evident that this measurement procedure is able to saturate the bound (\ref{alphabound}), but we will see that it does. In the following, we will see how to choose $p$, $\bf c$ and $\bf c'$. Our constructed joint measurement will have the averages
\begin{eqnarray}
\overline{A_J}&=&p\langle{\bf c\cdot\hat{\sigma}}\rangle+(1-p)\langle{\bf c'\cdot\hat{\sigma}}\rangle \nonumber \\
\overline{A'_J}&=&p\langle{\bf c\cdot\hat{\sigma}}\rangle-(1-p)\langle{\bf c'\cdot\hat{\sigma}}\rangle .\end{eqnarray}
We require the joint measurement to satisfy
\begin{eqnarray}
\overline{A_J}&=&\alpha\langle{\bf a\cdot\hat{\sigma}}\rangle\nonumber\\
\overline{A'_J}&=&\alpha '\langle{\bf a'\cdot\hat{\sigma}}\rangle .
\end{eqnarray}
For this to be true for all possible states we must have 
\begin{eqnarray}
\label{ccond}
\bf c &=& {1\over{2p}}(\bf \alpha a + \alpha 'a')\nonumber\\
\bf c' &=& {1\over{2(1-p)}}(\bf\alpha a - \alpha 'a').
\end{eqnarray}
Since $\bf c$ and $\bf c'$ are required to be unit vectors, it follows that
\begin{eqnarray}
\label{pcond}
p&=&{1\over 2}|\alpha \bf a + \alpha ' a'|\nonumber\\
1-p&=&{1\over 2}|\alpha \bf a - \alpha ' a'|.
\end{eqnarray}
Eliminating $p$ gives
\begin{equation}
{\bf |\alpha a + \alpha ' a'| + |\alpha a - \alpha ' a'|}= 2,
\end{equation}
which means that our joint measurement reaches the equality in (\ref{alphabound}). It is therefore an optimal joint measurement. 

We can also make a connection with the previous section, where we were considering a singlet state. Looking more closely at the derivation leading to condition (\ref{alphabound}), particularly at the step in equation (\ref{ineq}), one sees that the bound is satisfied if and only if one of the probabilities $p(A_J=A'_J=B)$ and $p(A_J=A'_J=-B)$, and also one of the probabilities $p(A_J=-A'_J=B')$ and $p(A_J=-A'_J=-B')$ are equal to zero. Suppose we have an EPR pair of two spin-1/2 particles and choose to perform the joint measurement on one of the particles by measuring spin {\it either} along $\bf c$ {\it or} along $\bf c'$. On the other particle, we measure either $\bf b$ or $\bf b'$. If now both $\bf b$ and $\bf c$ are chosen parallel to $\bf\alpha a+\alpha ' a'$, and $\bf b'$ and $\bf c'$ parallel to $\bf\alpha a-\alpha ' a'$, measurement results will be correlated in the following way. 
Obtaining $C=+1$ means that $A_J=B_J=+1$; this is also perfectly correlated with obtaining $B=+1$. Therefore  the probability $p(A_J=A'_J=-B)$ will be zero. If instead we choose $\bf b$ and $\bf c$ {\it antiparallel} to each other, $p(A_J=A'_J=B)$ would be zero instead. 
Similarly, one finds that either $p(A_J=-A'_J=B')$ or $p(A_J=A'_J=-B')$ will be zero.  The joint measurement we have considered will therefore satisfy the bound (\ref{alphabound}).

We have just seen that one way to realise the joint measurement of $\bf a\cdot\hat{\sigma}$ and $\bf a'\cdot\hat{\sigma}$ is to measure spin {\it either} along $\alpha \bf a+\alpha 'a'$, with probability $p
={1\over 2}|\alpha \bf a + \alpha ' a'|$, {\it or} along $\alpha \bf a-\alpha 'a'$, with probability $1-p={1\over 2}|\alpha \bf a - \alpha ' a'|$. This is only one of (infinitely) many ways to realise the joint measurement. Another possibility is to start with the quantum system we want to measure, and then couple this quantum system, in a suitable way, to an auxiliary system in a known state. This is then followed by measurements of spin for both systems, along suitable directions. A third approach would be to clone the original quantum system and then to measure one observable on each of the clones. Perfect cloning is not possible, and any attempt will result in extra noise in the joint measurement. Any realisation of a joint measurement must satisfy the bound we have derived. A universal cloner, for example, will take a state $\rho = 1/2(\mathbf{1}+\bf m\cdot\hat{\sigma})$ to two identical copies with $\rho' = 1/2(\mathbf{1}+\eta\bf m\cdot\hat{\sigma})$, where $\eta\le2/3$, and $\eta = 2/3$  for an optimal universal cloner \cite{buzekhillery}. If we measure one observable on each clone, we will obtain
 $\alpha = \alpha' =\eta \le 2/3$ for any directions $\bf a$ and $\bf a'$. Therefore the joint measurement using universal cloning is not optimal. Even when $\bf a$ and $\bf a'$ are orthogonal, and  condition (\ref{alphabound}) is as strict as it can be, it gives $\alpha = \alpha' \le 1/\sqrt{2}$  if $\alpha$ and $\alpha'$ are equal, and $2/3<1/\sqrt{2}$. The reason why universal cloning does not result in an optimal joint measurement is probably that the universal cloner clones any state equally well. In our case, it would be better to clone the basis states $|\pm\rangle_a$ and $|\pm\rangle_b$, and states close to these, as well as possible. We will return to this topic elsewhere.

\section{Generalised measurement operators}
\label{secjointop}

In this section, we will look at the generalised measurement operators of the joint spin measurement. 
The measurement operators obtained here will describe {\it any} realisation of our joint quantum measurement. No previous knowledge of generalised measurements will be required to follow our treatment, only basic quantum mechanics.

A von Neumann measurement is a projection onto the eigenstates of the measured  observable. When measuring the spin observable $\bf a\cdot\hat{\sigma}$, the projectors onto the eigenstates are given by
\begin{equation}
\label{ameas}
\Pi_\pm^a={1\over 2}(\mathbf{1}\pm\bf a\cdot\hat{\sigma}).
\end{equation}
Joint measurements of non-commuting observables cannot directly be described as a projection onto eigenstates in this way, because the observables do not share eigenstates. However, they can be described as generalised measurements, so-called POM (probability operator measure) or POVM (positive operator-valued measure) strategies \cite{hel, nch}. Generalised measurements allow us to describe any measurement that can be performed within the limits of quantum mechanics.
In fact, due to imperfect detectors, noise etc., any experimentally realised measurement is usually a generalised measurement rather than a projective measurement.

First we will describe the realisation of joint measurement which was discussed in the previous section.
In analogy with equation (\ref{ameas}), a measurement along $\bf c$ with a probability $p$ can be associated with the measurement operators
\begin{equation}
\Pi_\pm^c={p\over 2}({\mathbf{1}}\pm\bf c\cdot\hat{\sigma}),
\end{equation}
and similarly for a measurement of $\bf c'$ with the probability $1-p$. Using this fact, we find that
the joint measurement in section \ref{secjointmeas}, with four results, is described by the four measurement operators
\begin{eqnarray}
\label{jointoper}
\Pi^{aa'}_{++}&=&{1\over 4}|\alpha {\bf a + \alpha ' a'}| \mathbf{1}+ {1\over 4}(\alpha \bf a + \alpha ' a')\cdot\hat{\sigma}\nonumber\\
\Pi^{aa'}_{--}&=&{1\over 4}|\alpha {\bf a + \alpha ' a'}|\mathbf{1}- {1\over 4}(\alpha \bf a + \alpha ' a')\cdot\hat{\sigma}\nonumber\\
\Pi^{aa'}_{+-}&=&{1\over 4}|\alpha {\bf a - \alpha ' a'}|\mathbf{1}+ {1\over 4}(\alpha \bf a - \alpha ' a')\cdot\hat{\sigma}\nonumber\\
\Pi^{aa'}_{-+}&=&{1\over 4}|\alpha {\bf a - \alpha ' a'}|\mathbf{1}- {1\over 4}(\alpha \bf a - \alpha ' a')\cdot\hat{\sigma},
\end {eqnarray}
where we have used expressions (\ref{ccond}) and (\ref{pcond}) for $p, 1-p, \mathbf c$ and $\mathbf c'$.
The measurement operators $\Pi_i$ for  a generalised measurement do not have to be pure state projectors, but they do have to obey certain conditions. In analogy with projective measurements, the probability $p_i$ to obtain result $i$ is given by $Tr\{\Pi_i\rho\}$ for any measured state $\rho$. Because probabilities have to be nonnegative, all eigenvalues of $\Pi_i$ have to be greater than or equal to zero. Also, the measurement operators have to sum up to the identity operator, $\sum_i\Pi_i=\mathbf{1}$, corresponding to the fact that the sum of the probabilities $p_i$ is equal to 1. It is easy to check that the measurement operators (\ref{jointoper}) satify these conditions.

We can form the marginal measurement operators
\begin{eqnarray}
\label{margoper}
\Pi^a_+&=&\Pi^{aa'}_{++}+\Pi^{aa'}_{+-}={1\over 2}(\mathbf{1}+\bf \alpha a\cdot\hat{\sigma})\nonumber\\
\Pi^a_-&=&\Pi^{aa'}_{-+}+\Pi^{aa'}_{--}={1\over 2}(\mathbf{1}-\bf \alpha a\cdot\hat{\sigma})
\end{eqnarray}
describing the unsharp measurement of $\bf a\cdot\hat{\sigma}$, when measured jointly with $\bf a'\cdot\hat{\sigma}$. The measurement operators $\Pi^{a'}_\pm$ are obtained in an analoguous way. The amount of smearing depends on $|\alpha|$; the smaller it is, the more smeared the observable is. A sharp measurement of $\bf a\cdot\hat{\sigma}$ has $|\alpha|=1$, giving the measurement operators in equation (\ref{ameas}).

The measurement operators in equation (\ref{jointoper}) describe the optimal joint measurement. A more general choice of measurement operators, describing a joint measurement which does not have to be optimal \cite{busch}, is 
\begin{eqnarray}
\label{jointoper2}
\Pi^{aa'}_{++}&=&{1\over 4}(1+\alpha {\bf a \cdot \alpha ' a'}) \mathbf{1}+ {1\over 4}(\alpha \bf a + \alpha ' a')\cdot\hat{\sigma}\nonumber\\
\Pi^{aa'}_{--}&=&{1\over 4}(1+\alpha {\bf a \cdot\alpha ' a'})\mathbf{1}-{1\over 4}
(\alpha \bf a + \alpha ' a')\cdot\hat{\sigma}\nonumber\\
\Pi^{aa'}_{+-}&=&{1\over 4}(1-\alpha {\bf a \cdot\alpha ' a'})\mathbf{1}+ {1\over 4}
(\alpha \bf a - \alpha ' a')\cdot\hat{\sigma}\nonumber\\
\Pi^{aa'}_{-+}&=&{1\over 4}(1-\alpha {\bf a \cdot\alpha ' a'})\mathbf{1}- {1\over 4}
(\alpha \bf a - \alpha ' a')\cdot\hat{\sigma}.
\end {eqnarray}
It is easy to verify that these measurement operators also give the marginal measurement operators given in equation (\ref{margoper}). For the operators (\ref{jointoper2}) to describe a valid measurement, their eigenvalues have to be greater than or equal to zero. This means that the length of the vector $\alpha\bf a+\alpha 'a'$ has to be smaller than or equal to $1+\alpha\bf a\cdot\alpha 'a'$, and similarly for $\alpha\bf a-\alpha 'a'$ and $1-\alpha\bf a\cdot\alpha 'a'$. This condition turns out to be equivalent to the bound (\ref{alphabound}), which any joint measurement has to satisfy. When equality holds in condition (\ref{alphabound}), the operators in equation (\ref{jointoper2}) become identical to those in equation (\ref{jointoper}). 

\section{Uncertainty relation}
\label{secunc}

The variances of the jointly measured observables will exceed those found when the observables are measured separately. As we now will show, the bound (\ref{alphabound}) on the sharpness of the joint measurement can be directly related to an uncertainty relation for the jointly measured observables.

Squaring expression (\ref{alphabound}), and noting that $|\alpha{\bf a}\pm\alpha ' {\bf a}'|^2=\alpha^2+\alpha ^{\prime 2}\pm 2\alpha\alpha {\bf 'a\cdot a'}$, we obtain
\begin{equation}
{|\alpha\bf a+\alpha ' a'||\alpha\bf a-\alpha ' a'|}\leq 2-\alpha^2-\alpha^{\prime 2}.
\end{equation}
Squaring this expression once more, and cancelling terms on both sides, we obtain
\begin{equation}
\alpha^2+\alpha^{\prime 2}-\alpha^2\alpha^{\prime 2}({\bf a\cdot a'})^2\leq 1.
\end{equation}
This expression is already a useful way of expressing the bound on $\alpha$ and $\alpha '$. Denoting $\bf a\cdot a'$ by $\cos\theta$, we may also write it in a product form,
\begin{equation}
\label{alphaunc}
{(1-\alpha^2)(1-\alpha^{\prime 2})\over {\alpha^2\alpha^{\prime 2}}}\geq\sin^2\theta .
\end{equation}

The uncertainty in the joint measurement arises from two sources, the intrinsic uncertainty in the quantum observables, and the fact that they are measured jointly.
The variance in the joint measurement, squared and multiplied with $\alpha^{-2}$, can be written
\begin{eqnarray}
\Delta^2 A_J /\alpha^2&=& (1-\alpha^2\langle{\bf\hat{A}}\rangle^2)/\alpha^2 \nonumber\\
&=&(1-\alpha^2)/\alpha^2 + 1-\langle{\bf\hat{A}}\rangle^2,
\end{eqnarray}
and similarly for $\Delta^2 A'_J$.
Here $1-\langle{\bf\hat{A}}\rangle^2$ is the ``bare" variance of $\bf\hat{A}$, when measured on its own, and similarly for $\bf\hat{A'}$. The quantities $(1-\alpha^2)/\alpha^2$ and $(1-\alpha'^2)/\alpha'^2$ are seen to be contributions coming purely from the fact that the measurement is a joint measurement. A lower bound on their product is given by (\ref{alphaunc}), which is now understood to be an uncertainty relation giving a lower bound on the uncertainty associated purely with the fact that $\bf\hat{A}$ and $\bf\hat{A'}$ are quantum observables which are measured {\it jointly}. This bound is tight, meaning that there is always a measurement such that equality can be reached, and holds only for spin-1/2 particles. Furthermore, it does not depend on the measured state at all, only on the measured quantum observables. This is in contrast to the Heisenberg-Scr\"odinger-Robertson uncertainty relation \cite{heis, rob, schroe} for the variances for the separately measured observables. For spin-1/2 particles, this gives a bound which is {\it dependent} on the measured state, and depending on the state, the resulting uncertainty bound may not be tight. The Heisenberg-Schr\"odinger-Robertson  uncertainty relation 
for the product of the ``bare" variances is 
\begin{eqnarray}
\label{heisunc}
\Delta^2A\Delta^2A'&=&(1-\langle{\bf\hat{A}}\rangle^2)(1-\langle{\bf\hat{A'}}\rangle^2)
\geq{1\over 4}|\langle[ {\bf\hat{A},\bf\hat{A'}}]\rangle|^2\nonumber\\
&=& |\langle (\bf a \times a' )\cdot\hat{\sigma}\rangle|^2
=\sin^2\theta|\langle{\bf a}_\perp\cdot\hat{\sigma}\rangle |^2,
\end{eqnarray}
where $\bf a_\perp$ is perpendicular to both $\bf a$ and $\bf a'$. Using this together with (\ref{alphaunc}), it is possible to obtain a bound on the total uncertainty product for the joint measurement as
\begin{equation}
\label{totuncrel}
{\Delta^2 A_J\Delta^2 A'_J\over{\alpha^2\alpha'^2}}
\geq 
\sin^2\theta(1+|\langle{\bf a}_\perp\cdot\hat{\sigma}\rangle |)^2.
\end{equation}
This relation is valid only for joint spin measurements, and is stronger than 
the uncertainty relation for the variances of any two jointly measured observables derived by Arthurs and Goodman \cite{artgood}, which is
\begin{equation}
\label{totuncrelgen}
{\Delta^2 A_J\Delta^2 A'_J/( {\alpha^2\alpha'^2}})\geq |\langle[{\bf\hat{A},\bf\hat{A'}}]\rangle|^2 
= 4 \sin^2\theta |\langle{\bf a}_\perp\cdot\hat{\sigma}\rangle |^2 .
\end{equation}
This raises the question whether it is possible, in analogy with equation (\ref{totuncrel}), to derive a  tighter bound on joint measurements, which would hold in general and not only for spin measurements.

In deriving both joint measurement uncertainty relations above, we used the Heisenberg-Schr\"odinger-Robertson uncertainty relation, which is not always tight.
Schr\"odinger \cite{schroe} showed that a tighter uncertainty relation can be obtained,
\begin{eqnarray}
\label{commcorr}
&&\Delta^2A\Delta^2A'=(1-\langle{\bf\hat{A}}\rangle^2)(1-\langle{\bf\hat{A'}}\rangle^2)  \\
&&\geq {1\over 4}|\langle[ {\bf\hat{A},\bf\hat{A'}}]\rangle|^2+
{1\over 4}(\langle{{\bf\hat{A}\hat{A'}+\hat{A'}\hat{A}\rangle}-2\langle\bf\hat{A}\rangle\langle\hat{A'}}\rangle)^2,\nonumber
\end{eqnarray}
where the second term is  a covariance term. If relation (\ref{commcorr}) is used instead of  relation (\ref{heisunc}), one obtains a tighter bound on the joint measurement as well. The resulting expression for spin measurements, however, is not as nice and simple as equation (\ref{totuncrel}).

Once again we can make a connection to Cirel'son's inequality. Translating condition (\ref{cirelalpha}), which arose from the Cirel'son inequality, into a product bound, gives
\begin{equation}
{(2-\alpha^2)(2-\alpha^{\prime 2})\over {\alpha^2\alpha^{\prime 2}}}\geq\sin^2\theta .
\end{equation}
As it should, this condition is immediately seen to place no restriction that $|\alpha|$ and $|\alpha'|$ has to be less than 1.
Interestingly enough, Cirel'son's inequality can also be derived by considering uncertainty relations for suitably chosen new observables \cite{tonystevecirel}.

\section{Discussion and conclusions}

We have shown how the bound on the sharpness of a joint spin measurement may be obtained using a locality argument. Apart from operational locality, the derivation only requires the existence of a joint probability  distribution for the two measured spin components for a single spin 1/2 particle. This distribution must necessarily exist for any  joint measurement, whether quantum mechanical or not. The resulting measurement bound, in the form of Eq. (\ref{belleq}), very much resembles a CHSH Bell inequality. This inequality must be satisfied by the correlation functions for the joint quantum measurement. Our derivation can also be understood as making a connection between the CHSH Bell inequality and the bound on joint spin measurements. It has in fact been shown that the Bell inequalities follow from the existence of a joint probability distribution of all triples of observables occurring in the  Bell inequality, producing the correct marginal probability distributions, and also conversely, that if the Bell inequalities hold, then such a joint probability distribution exists  \cite{fine}. Ekert has also shown that Bell's theorem can be used to test for eavesdropping in quantum cryptography \cite{ekert}, the security of which relies on the impossibility of performing ideal joint measurements. 

In this paper we have used the principle of operational locality much like a physical law, in the same fashion as energy conservation may be used for shorter and more elegant solutions of physical problems. The derivation of the bound on joint spin measurements is not an isolated example. In an earlier paper, we demonstrated how to derive a bound on how well two non-orthogonal states may be distinguished from each other using a locality argument \cite{bar}. Another example is the bound on symmetric cloning \cite{gisin}. It is interesting to ask exactly how much of the framework of quantum mechanics follows from the assumption of operational locality.
It seems that, in order to obtain quantum mechanics, other assumptions must be added as well. For example, complete positivity does not appear to follow from the assumption of operational locality, but must be added as an axiom on its own \cite{dagmar}. To appreciate this, consider two separated, possibly entangled physical systems. Partially transposing the density matrix of one of the systems is known to not to be a completely positive operation \cite{nch}. Nevertheless, the partial transpose does not in any way alter the reduced density matrix of the other system, and so must be considered to satisfy the principle of operational locality.
 
An example of a realisation of the joint measurement was also given. It turns out that by measuring spin along one or the other of two well chosen directions $\bf c$ and $\bf c'$, an optimal joint measurement along $\bf a$ and $\bf a'$ can be realised. One point should perhaps be clarified. Suppose that we add another arm to our setup so that it looks like a setup used to test Bell's inequality \cite{alain}. In the first arm, a measurement along either $\bf c$ or $\bf c'$ is made, in the second arm, along either $\bf b$ or $\bf b'$. Now, if we use the measurement results along $\bf c$ and $\bf c'$ to construct a joint measurement along $\bf a$ and $\bf a'$, then the Bell-like condition for the joint measurement, given in equation (\ref{belleq}) involving the $\bf a, a', b$ and $\bf b'$ directions, will be satisfied.  However, if we use the measurement results to construct correlation functions for the $\bf b$, $\bf b'$, $\bf c$ and $\bf c'$ directions, and quantum mechanics is valid, our results may violate a {\it different} Bell inequality involving the $\bf b, b', c$ and $\bf c'$ directions. 
In the realisation of the quantum joint measurement it does not matter if we use a beam splitter or an active ``classical" switch to make the choice between $\bf c$ and $\bf c'$. However, in an experiment designed to test local realism, there is a difference between actively deciding which direction to measure along and letting the beam splitter determine the observable to be measured. The crucial point is that the choice should not  depend on any hidden variables.

We have also discussed uncertainty relations for joint measurements of spin. The joint measurement bound can be rewritten as a bound on the product of the necessary increases in the variances. In contrast with the familiar Heisenberg-Schr\"odinger-Robertson uncertainty relation,  our bound is tight, and independent of the measured state. This means that,  for any given quantum state, there always is a measurement which will saturate the bound. The increase in uncertainty, resulting from the fact that we perform a joint measurement of spin along two directions, does not depend on the measured state. 
The  bound derived here holds only for joint measurements of spin-1/2. This raises the question whether it is possible to derive similar tight bounds for joint measurements of observables other than spin-1/2 \cite{son}. Also, one could consider joint measurements of more than two observables, such as spin along any three linearly independent directions. This could give rise to uncertainty-relation-like conditions for more than two jointly measured observables. To our knowledge, the question of uncertainty relations for more than two  observables was first raised by Robertson, who gave a bound for the product of the variances for an even number of 
observables  \cite{rob2}. 

\acknowledgments
We want to thank David Pegg for useful discussions.
Financial support from the Carnegie Trust, the Royal Society of Edinburgh, the Scottish Executive Education and Lifelong Learning Department, the EU Marie Curie programme, project number HPMF-CT-2000-00933, the Dorothy Hodgkin Fellowship scheme of the Royal Society of London, and from the INTAS network is gratefully acknowledged.

\end{document}